\documentclass[traditabstract]{aa}
\usepackage{txfonts}
\usepackage{graphicx}
\usepackage{natbib}
\usepackage{longtable}
\bibpunct{(}{)}{;}{a}{}{,}

\begin{document}

\title{Orbital period changes and the higher-order multiplicity
  fraction amongst SuperWASP eclipsing binaries\thanks{Tables A.1 to A.6 are available in electronic form at the CDS via anonymous ftp to cdsarc.u-strasb.fr (130.79.128.5) or via http://cdsweb.u-strasbg.fr/cgi-bin/qcat?J/A+A/}}
\author{M.~E.~Lohr\inst{\ref{inst1}}\and
  A.~J.~Norton\inst{\ref{inst1}}\and S.~G.~Payne\inst{\ref{inst1}}\and
  R.~G.~West\inst{\ref{inst2}}\and P.~J.~Wheatley\inst{\ref{inst2}}}
\institute{Department of Physical Sciences, The Open University,
  Walton Hall, Milton Keynes MK7\,6AA, UK\\
  \email{Marcus.Lohr@open.ac.uk}\label{inst1}\and Department of
  Physics, University of Warwick, Coventry CV4\,7AL,
  UK\label{inst2}} \date{Received 26 January 2015 / Accepted 29 April 2015}

\abstract {Orbital period changes of binary stars may be caused by the
  presence of a third massive body in the system.  Here we have
  searched the archive of the Wide Angle Search for Planets
  (SuperWASP) project for evidence of period variations in 13927
  eclipsing binary candidates.  Sinusoidal period changes, strongly
  suggestive of third bodies, were detected in 2\% of cases; however,
  linear period changes were observed in a further 22\% of systems.
  We argue on distributional grounds that the majority of these
  apparently linear changes are likely to reflect longer-term
  sinusoidal period variations caused by third bodies, and thus
  estimate a higher-order multiplicity fraction of 24\% for SuperWASP
  binaries, in good agreement with other recent figures for the
  fraction of triple systems amongst binary stars in general.}

\titlerunning{Period changes and multiplicity in SuperWASP binaries}
\authorrunning{M.~E.~Lohr et al.}

\maketitle
\section{Introduction}

The importance of multiplicity for an understanding of stellar
evolution is hard to overestimate.  Single stars now appear to be in a
minority, especially at higher masses (the recent review of
\citet{duchene} indicated a multiplicity fraction for
intermediate-mass stars $\ge$50\%, rising to $\ge$80\% for the most
massive stars), and binary interactions are probably responsible for
creating several different types of supernovae, novae and unusual star
types such as blue stragglers.  Amongst binaries, a significant
proportion appear to be part of higher-order multiple systems
(\citet{tokovinin1,tokovinin2} estimated 29\% for F and G dwarfs in a
distance-limited sample), and such triples, quadruples etc. also have
much to tell us about the formation and stability of stellar systems
(see e.g. \citet{michaely,naoz}).

Higher-order multiple star systems can be detected by a range of
methods including direct resolution, radial velocity and proper motion
analysis.  In several recent papers, we have used archive light curves
from the Wide Angle Search for Planets (SuperWASP) \citep{pollacco} to
study eclipsing binaries, benefiting from its high-cadence,
long-baseline time-domain photometry.  With such data, additional
eclipses in a binary light curve may occasionally reveal further
bodies in a system, as with the doubly-eclipsing quintuple system
reported in \citet{lohr13} and followed up by \citet{koo}.  However, a
more widely-applicable technique will be the detection of eclipse
timing variations producing an approximately sinusoidal curve in an
O$-$C (observed minus calculated) diagram, as in \citet{lohr13b},
where a triple system containing an M+M contact binary was proposed on
such evidence, and later confirmed by \citet{koen}; in \citet{lohr14b}
also, period variations arguably produced by circumbinary planets were
detected in several post-common-envelope binaries.  This approach was
used by \citet{rappaport} to search for candidate triples in the
Kepler eclipsing binary catalogue.  Here, we use a preliminary
catalogue of candidate SuperWASP eclipsing binaries to search for
orbital period changes potentially indicative of third bodies.  These
statistics can then be used to suggest a lower limit to the
higher-order multiplicity fraction of SuperWASP stars.

\section{Method}

A provisional catalogue of SuperWASP eclipsing binary candidates was
produced by \citet{payne}, using a neural net classification method
for all objects listed in the database with periods found by the
method described in \citet{norton07,norton}.  The catalogue contained
2875 objects classified as EW-type (probable contact) binaries, 5226
EB-type (light curve resembling $\beta$~Lyrae), 5826 EA-type (light
curve resembling Algol=$\beta$~Persei) exhibiting two eclipses per
cycle, and 7056 potential EA-type systems in which only a single
eclipse was visible.  Owing to the large number of false positives
expected in the last group, only the first three groups of sources
were considered here for further analysis.

13927 light curves were downloaded from the SuperWASP archive, and a
form of the custom IDL code described in \citet{lohr14b}, modified for
large numbers, was run on them.  This checked and refined the orbital
period associated with each object identifier, searching within a
range centered on the catalogue period (itself derived from the
archive database); produced a phase-folded light curve and mean
fitting curve (with 100 bins); generated O$-$C, amplitude change and
absolute flux change diagrams; and determined a rate of period change
where this was supported by the O$-$C diagram.  The output for each
identifier was an image file allowing visual checking of the light
curve, O$-$C and other diagrams; and a log file line summarizing key
statistics such as number of data points in the light curve, mean
flux, orbital period and evidence for period change.

\begin{figure}
\resizebox{\hsize}{!}{\includegraphics{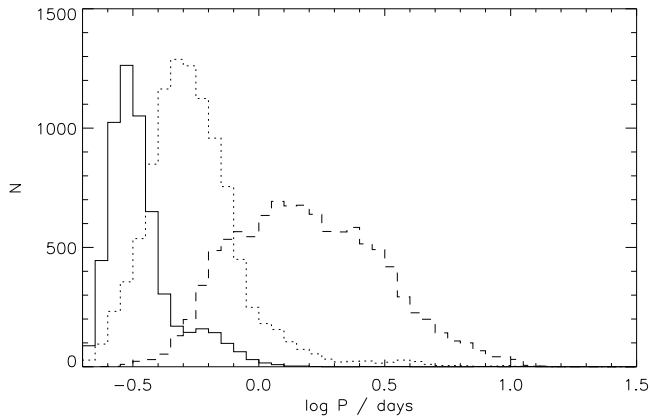}}
\caption{Period distributions of SuperWASP eclipsing binary candidates
  after preliminary quality checks.  EW-type objects are shown with a
  solid line (cf. \citet{paczynski} EC group); EBs with a dotted line
  (cf. \citeauthor{paczynski}'s ESD group); EAs with a dashed line
  (cf. \citeauthor{paczynski}'s ED group).}
\label{allbinshist}
\end{figure}

The log file revealed a number of objects clustered near particular
periods: 1/4, 1/3, 2/5 and 2/3 of a sidereal day in particular.
Visual checks of these objects (mainly on the EA list) confirmed that
they were spurious periodic variables, with the variability probably
resulting from temperature-related instrumental effects occurring
daily e.g. peaks or troughs in the light curve at the start of each
night's observations.  These identifiers were removed from the list.
For some objects, no significant period could be found in the given
range, or no O$-$C diagram could be constructed (due to insufficient
successful fits to nightly observations), and these objects were also
removed from further consideration.  These preliminary checks left
2844 EW-type, 5073 EB-type and 5323 EA-type objects.

Fig.~\ref{allbinshist} shows their period distribution by type; a
broadly comparable collection from the ASAS database is given in
\citet{paczynski} Fig.~6.  Although our collection contains
proportionally far fewer contact-type binaries (or perhaps far more of
the other two types), the peaks and ranges of the distributions are
very similar: EWs peak around $P=0.3$--0.35~d and tail off above about
1~d; EBs peak around $P=0.5$~d and are very rare beyond 2~d; both
types drop off sharply in period below the ``short period cut-off'' at
around $P=0.20$~d (see \citet{lohr,lohr13} for more detail on the
period distribution of SuperWASP eclipsing binaries in this region).
EAs have a broader peak around $P=1$--3~d, and are very rare below
$P=0.3$~d and above $P=10$~d.  Catalogues of specific types of
eclipsing binaries, such as classical Algols (see
\citet{budding,ibanoglu} and especially \citet{vanrensbergen} Fig.~5),
exhibit greater numbers at longer periods; however, these are less
likely to be detected reliably in SuperWASP data due to time-sampling
limitations or because they are brighter than $V\sim9$ mag.

\begin{figure}
\resizebox{\hsize}{!}{\includegraphics{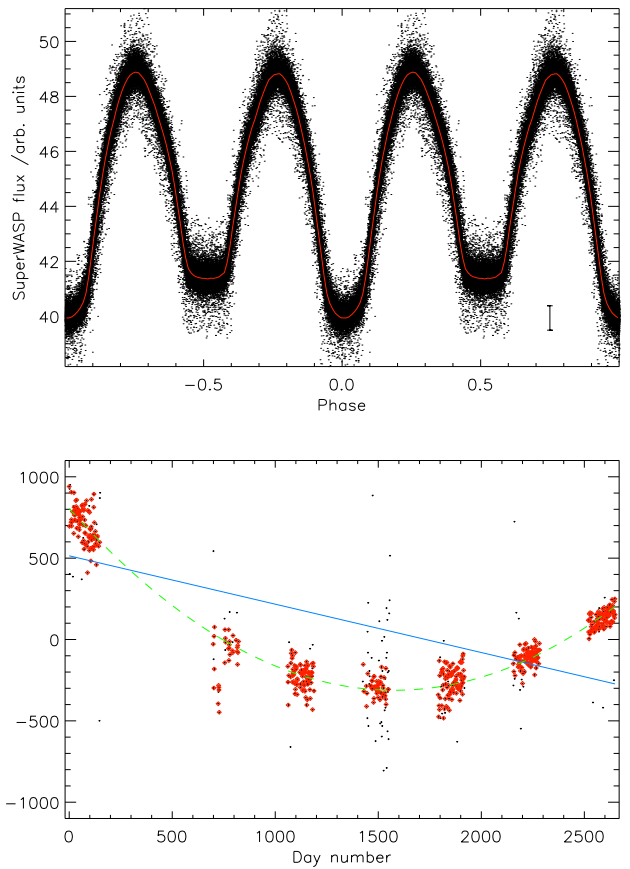}}
\caption{Top: Light curve for J171747 (automatically classified as
  EW-type) folded at $P=38649.224\pm0.006$~s, with binned mean curve
  overplotted in red (colour version in online edition only).  A
  representative error bar for a single observation is shown in the
  lower corner.  Bottom: O$-$C diagram for J171747 spanning eight
  years.  Red (larger) points were automatically selected for period
  change determination; black (smaller) points were excluded as
  outliers (a few additional more extreme outliers fall outside the
  bounds of the plot).  Blue solid line shows best linear fit (reduced
  $\chi^2=13.58$, 548 degrees of freedom); green dashed line shows
  best quadratic fit ($\chi^2=1.03$, 547~d.f.), strongly supporting a
  secular period increase, with rate $0.1466\pm
  0.0018$~s~yr\textsuperscript{-1}.}
\label{J171747}
\end{figure}

For the remaining objects, where period change had been found by the
code, the ratio between best linear and quadratic fit reduced $\chi^2$
values for the O$-$C diagram was used to select a sample for visual
checking.  All output files were checked down to a ratio of 1.25,
below which it was generally difficult to judge the classification
reliably by eye; tests applied to eclipsing post-common-envelope
binaries in \citet{lohr14b} had also indicated that ratios below 1.05
did not generally indicate statistically significant period change.
This meant that 679 EW-type, 436 EB-type and 806 EA-type objects were
checked visually, and assigned a classification: plausible quadratic
variation in the O$-$C diagram (supporting secular period change);
plausible sinusoidal variation (supporting alternating period
increases and decreases); no apparent period change (usually due to
erratic time sampling misleading the program's fitting algorithm);
erroneous period found (usually due to the original input period being
significantly wrong); or unclear (usually when the time sampling was
very sparse or the time basis very limited).

\section{Results}

\begin{figure}
\resizebox{\hsize}{!}{\includegraphics{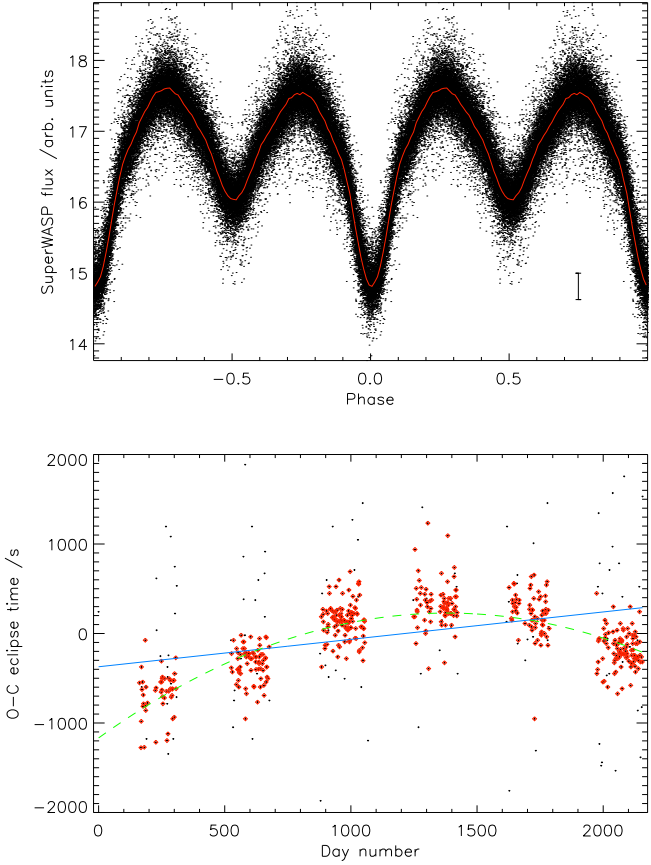}}
\caption{Top: Light curve for J064024 (automatically classified as
  EB-type) folded at $P=44851.164\pm0.007$~s.  Bottom: O$-$C diagram
  for J064024 covering six years.  Blue solid line shows best linear
  fit ($\chi^2=2.96$, 466~d.f.); green dashed line shows best
  quadratic fit ($\chi^2=1.01$, 465~d.f.), strongly supporting a
  secular period decrease, with rate $-0.277\pm
  0.009$~s~yr\textsuperscript{-1}.}
\label{J064024}
\end{figure}

\begin{figure}
\resizebox{\hsize}{!}{\includegraphics{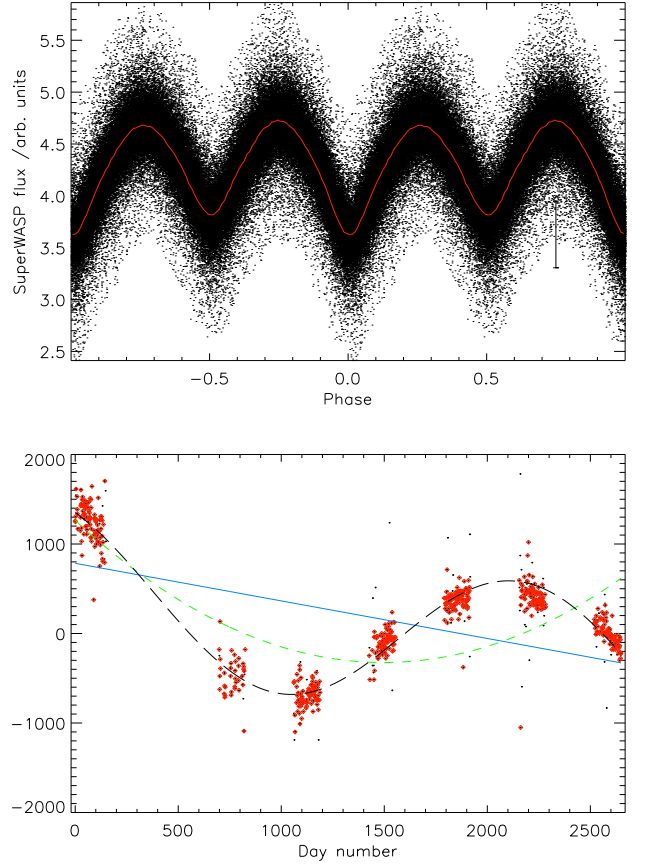}}
\caption{Top: Light curve for J165649 (automatically classified as
  EB-type) folded at $P=23775.304\pm0.003$~s.  The short period and
  light curve shape would probably support EW-type classification
  instead.  Bottom: O$-$C diagram for J165649 spanning eight years.
  Blue solid line shows best linear fit ($\chi^2=4.51$, 582~d.f.);
  green dashed line shows best quadratic fit ($\chi^2=2.14$,
  581~d.f.), which are similarly poor matches to the data.  Black
  long-dashed line shows best sinusoidal fit ($\chi^2<1.00$ using same
  O$-$C uncertainties as for linear and quadratic fits, 580~d.f.),
  with semi-amplitude $870\pm9$~s and modulation period
  $2386\pm11$~d.}
\label{J165649}
\end{figure}

\begin{figure}
\resizebox{\hsize}{!}{\includegraphics{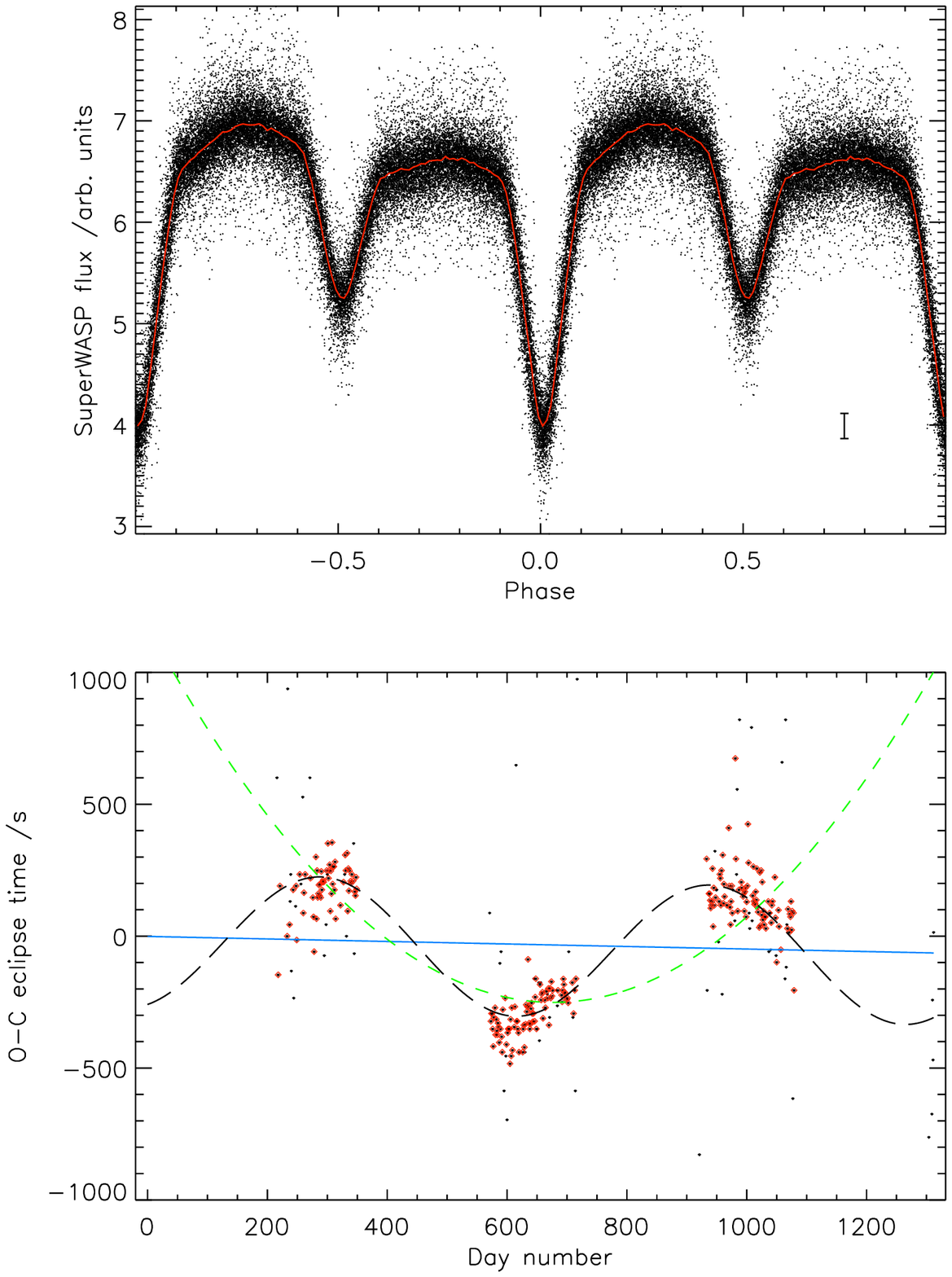}}
\caption{Top: Light curve for J161253 (automatically classified as
  EA-type) folded at $P=35162.816\pm0.007$~s.  Bottom: O$-$C diagram
  for J161253 covering just over three years.  Blue solid line shows
  best linear fit ($\chi^2=4.15$, 235~d.f.); green dashed line shows
  best quadratic fit ($\chi^2=1.10$, 234~d.f.).  Although a reasonable
  quadratic fit to the first three years of data has been achieved,
  the O$-$C trends within each year and the location of the partial
  data from year 4 on the diagram (small points around day 1300) lend
  more support to a sinusoidal variation.  Black dotted line shows
  best sinusoidal fit ($\chi^2<1.00$ using same O$-$C uncertainties as
  for linear and quadratic fits, 233~d.f.), with semi-amplitude
  $256\pm6$~s and modulation period $649\pm9$~d.}
\label{J161253}
\end{figure}

Period change was indicated by our code for 2305/2844 EW-type,
3227/5073 EB-type and 3076/5323 EA-type objects.  However, these
fractions cannot be taken at face value: large numbers of these
apparent changes involved very small differences between the best
linear fit and the best quadratic fit to the O$-$C diagram, which
would probably not have been statistically significant; many of the
EA-type objects exhibiting apparent period change also turned out to
have erroneous periods - usually those with very long periods which
were poorly sampled - and this could create the illusion of quadratic
period change.

480/679 EW-type visually-checked objects were classified as exhibiting
plausible period change, of which 388 showed quadratic and 92
sinusoidal behaviour (in the online appendix, Tables~\ref{ewquadtab}
and \ref{ewsinetab} give a complete list).  Fig.~\ref{J171747}
illustrates a clear case of period increase in this type.  Extending
this proportion to all the objects with linear-quadratic fit ratios
above 1.05, and adjusting the whole sample size to account for
expected numbers of erroneous periods and uncertain cases, we can
estimate that about 41\% of the EW-type objects are undergoing period
change (see Table~\ref{tabnumbers} for the full figures used in this
calculation).

167/436 EB-type objects were classified as exhibiting plausible period
change, 137 quadratic and 30 sinusoidal (Tables~\ref{ebquadtab} and
\ref{ebsinetab}).  Fig.~\ref{J064024} illustrates secular period
change in this type, while Fig.~\ref{J165649} shows very clear
sinusoidal variation (though this might be better described as an
EW-type system).  Scaling up the numbers as before, we can estimate
that about 19\% of the EB-type objects are undergoing period change.

189/806 EA-type objects were classified as exhibiting plausible period
change, 172 quadratic and 17 sinusoidal (Tables~\ref{eaquadtab} and
\ref{easinetab}).  Fig.~\ref{J161253} shows a probable case of
unusually short-term sinusoidal variation in this type.  Scaling up
the numbers as before, we can estimate that about 14\% of the EA-type
objects are undergoing period change.

\begin{figure}
\resizebox{\hsize}{!}{\includegraphics{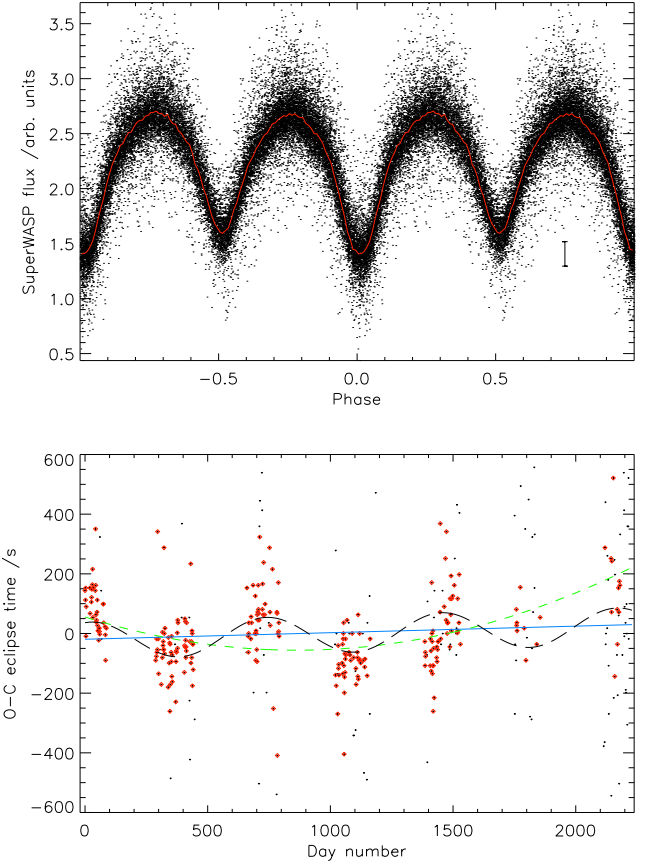}}
\caption{Top: Light curve for J142508 (automatically classified as
  EW-type) folded at $P=21568.141\pm0.002$~s.  Bottom: O$-$C diagram
  for J142508 spanning seven years.  Blue solid line shows best linear
  fit ($\chi^2=1.28$, 252~d.f.); green dashed line shows best
  quadratic fit ($\chi^2=1.01$, 251~d.f.).  Visual checks revealed
  that the apparently good quadratic fit relies in part on the
  widely-scattered poorer-quality data in the last two observing
  seasons, and appears inconsistent with the bulk of the
  closely-grouped observations in the third season (around days
  700--800); a sinusoidal fit was therefore preferred.  Black
  long-dashed line shows best sinusoidal fit ($\chi^2<1.00$ using same
  O$-$C uncertainties as for linear and quadratic fits, 250~d.f.),
  with semi-amplitude $62\pm7$~s and modulation period $710\pm50$~d.}
\label{J142508}
\end{figure}

Of the O$-$C diagrams judged to be exhibiting sinusoidal variation,
most ($\sim$70\% for EW-type; $\sim$90\% for EBs and EAs) show
approximately one complete cycle (as in Fig.~\ref{J165649}), with
near-equal proportions of such cases starting with period increase or
decrease i.e. there does not seem to be a detection bias in favour of
either.  Since sinusoidal fitting could not be reliably automated for
these (often noisy) data sets, best-fit modulation periods were not
obtained for the majority of objects, but visual estimation suggested
that the binaries with shorter orbital periods (mostly EW-type) tend
to possess shorter modulation periods.  This probably explains why
around 30\% of the sinusoidally-varying EWs show more than one
complete cycle during their time of observation by SuperWASP, while
only about 10\% of EBs and EAs do; Fig.~\ref{J142508} illustrates one
such case where three complete cycles have arguably been captured.

\begin{figure}
\resizebox{\hsize}{!}{\includegraphics{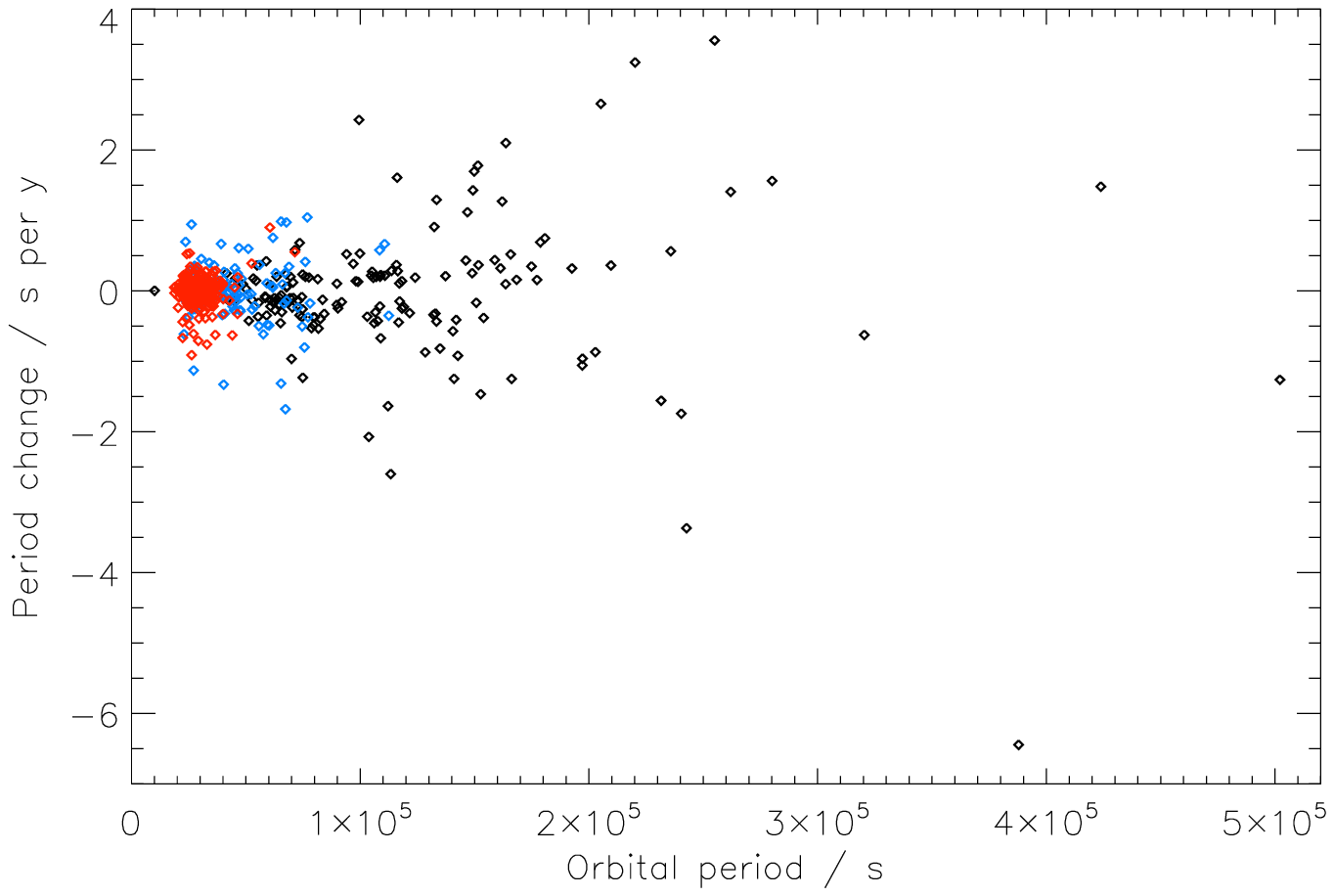}}
\caption{Period change measurements for 639 distinct sources
  exhibiting apparent quadratic variation in their O$-$C diagrams,
  plotted against their orbital periods.  EW-type binaries are shown
  in red, EB-types in blue, and EA-types in black.}
\label{perchgall}
\end{figure}

\begin{figure}
\resizebox{\hsize}{!}{\includegraphics{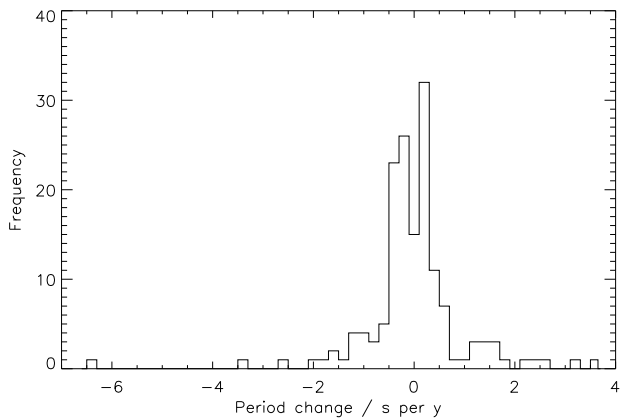}}
\caption{Period change distribution for EA-type eclipsing binaries.}
\label{eahist}
\end{figure}

Plotting (quadratic) period change measurements against periods
(Fig.~\ref{perchgall}) for the three types of eclipsing systems, we
may note that the most rapid changes are found in the long-period
EA-type binaries, while the short-period EWs all have period changes
below 1~s~yr\textsuperscript{-1} in magnitude.  The shortest
$P/\dot{P}$ timescale (i.e. a time to merger, if the period decrease
continued at this rate) is seen in EB/EW-type object J051927, at
$\sim$24000 years.

Nearly equal proportions of period increases (51.6\%) and decreases
(48.4\%) are found, without significant differences between the three
types.  Moreover, the period change distributions for the three types
(Figs.~\ref{eahist} to \ref{ewhist}) are very similar in shape: peaked
strongly at small values on each side of zero, with approximately
Gaussian tails at larger values, and a gap around zero itself, where
genuine period changes are very hard to detect.  A Kolmogorov-Smirnov
test of the positive values of each distribution against the negative
values does not support any significant difference between them
i.e. they are symmetric in a statistical sense.  However, when all
three distributions are considered simultaneously, the K-S test
provides near-significant support for the two sides being drawn from
different underlying distributions ($P$=0.07), and we may note that
the peak on the positive side is slightly higher than the peak on the
negative side in all three histograms.

\section{Discussion}

\begin{figure}
\resizebox{\hsize}{!}{\includegraphics{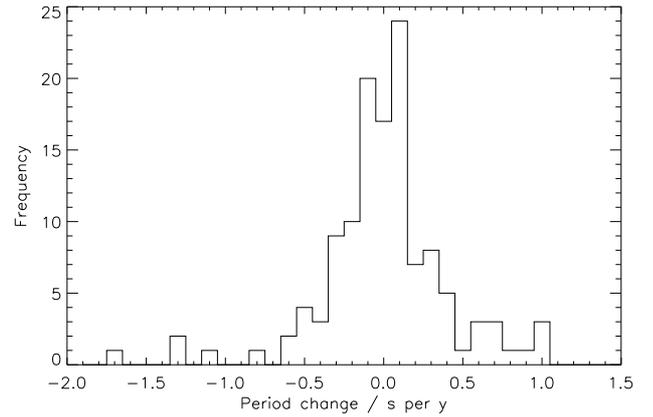}}
\caption{Period change distribution for EB-type eclipsing binaries.}
\label{ebhist}
\end{figure}

\begin{figure}
\resizebox{\hsize}{!}{\includegraphics{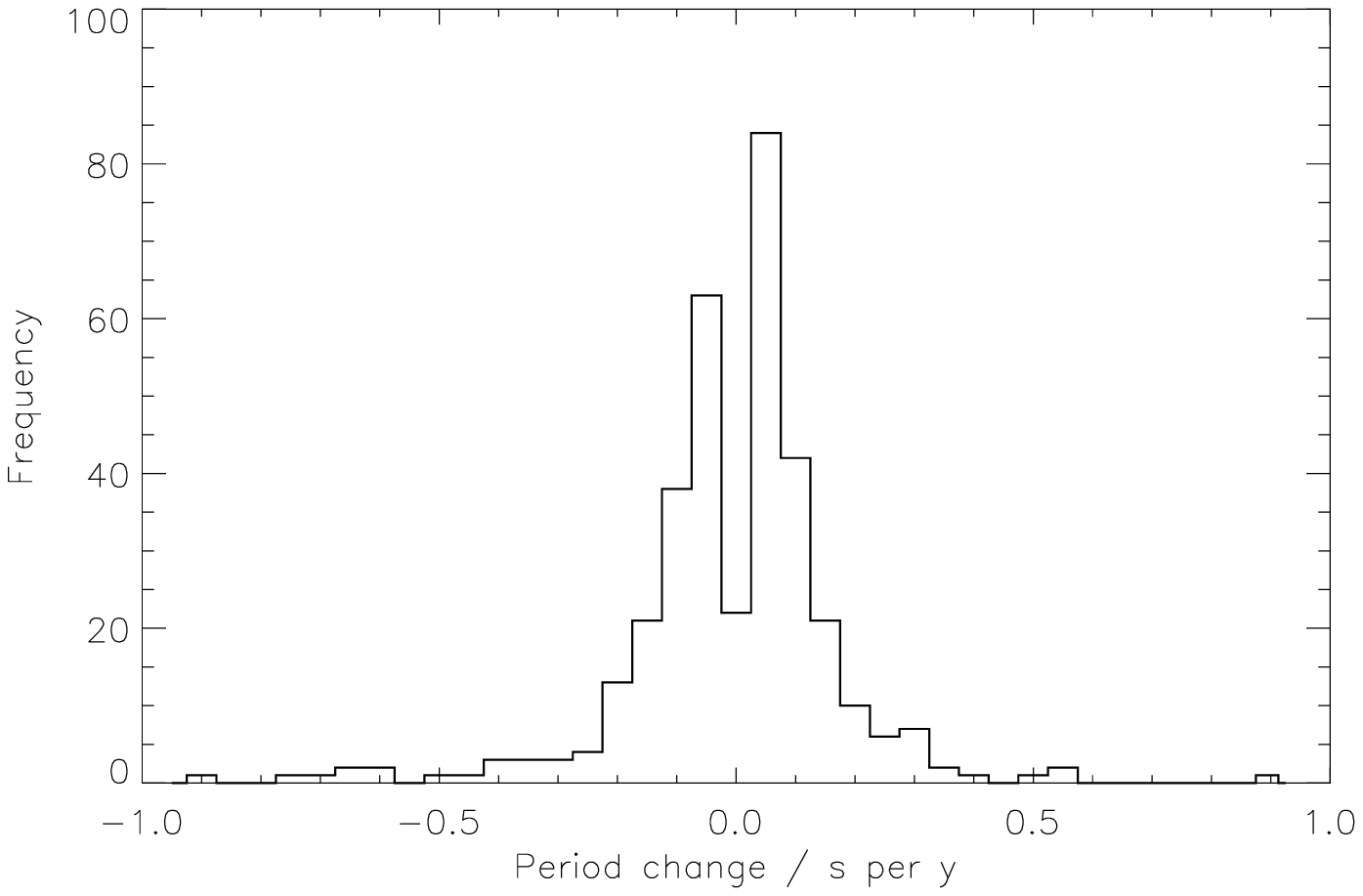}}
\caption{Period change distribution for EW-type eclipsing binaries.}
\label{ewhist}
\end{figure}

As was found in \citet{lohr13}, it is notable that the (quadratic)
period change distributions found here are all broadly symmetric, with
binary systems of all three light curve classes apparently as likely
to increase in orbital period as to decrease.  The three distributions
are also similar in shape, differing primarily in scale (the longer
the orbital period, the greater the typical rate of period change, so
that $\dot{P}/P$ is roughly constant).  This symmetry would not
necessarily be expected given the usual model of W~UMa-type contact
binaries forming ultimately from low-mass detached systems as they
move from wider to smaller separations, primarily due to magnetic
braking (e.g. \citet{hilditch,stepien12}); indeed, if there is a
slight asymmetry in Figs.~\ref{eahist} to \ref{ewhist}, it is in the
direction of period increase rather than decrease.  Moreover, such
evolution by magnetic braking would normally be expected to be two or
three orders of magnitude slower than the changes measured here:
\citet{eggleton} tentatively suggests a binary composed of Solar-type
stars might decrease in period from a few days to contact ($\sim0.3$
days) in something like the Hubble time.  More massive Algol-type
systems are currently modelled \citep{siess,deschamps} as undergoing
initial period decrease associated with mass transfer, followed by a
substantial period \emph{increase} after mass ratio reversal; however,
the majority of SuperWASP objects may be expected to be relatively
low-mass (the survey's magnitude limits mean that it mainly detects
sources in the local volume of the Milky Way, in which lower-mass
stars predominate).  In any case, mass-transfer evolution would not
produce a symmetrical period change distribution either.

An explanation might be found by considering the sinusoidal period
changes clearly seen in some of the O$-$C diagrams with long baselines
e.g. J165649.  In Fig.~\ref{J165649}, if we only had the first five
or six years of observations, the data would be well-fitted by a
quadratic opening upwards, and we should conclude that the system was
undergoing rapid period increase; conversely, if we only had the last
five years of data, the diagram would support a quadratic opening
downwards, and the system would appear to be undergoing steady period
decrease.  Given this, it seems plausible that many of the apparent
quadratic changes detected here would prove to be part of longer-term
sinusoidal variations if we continued to observe the systems.  If the
majority of period changes in our data set are actually short sections
of sinusoidal variations, this would neatly explain the symmetric
distributions seen in Figs.~\ref{eahist} to \ref{ewhist}, since the
sections would be equally likely to be drawn from any part of the
underlying sinusoid, implying equal numbers of apparent positive and
negative period changes, on average, for a large sample.

\begin{table*}
\caption{Object statistics used in multiplicity calculations}
\label{tabnumbers}
\centering
\begin{tabular}{l l l l l}
\hline\hline
 & & EW-type & EB-type & EA-type \\
\hline
1 & Total objects considered & 2844 & 5073 & 5323 \\
2 & Period change found by code & 2305 & 3227 & 3076 \\
3 & Objects with $\chi^2$ ratio $\ge$1.05\tablefootmark{a} & 1414 & 1395 & 1797 \\
4 & Objects without apparent period change\tablefootmark{b} & 1430 & 3678 & 3526 \\ 
5 & Objects checked visually & 679 & 436 & 806 \\
6 & Period change confirmed visually & 480 & 167 & 189 \\
7 & Period change rejected visually & 88 & 50 & 48 \\
8 & Wrong period detected visually & 24 & 33 & 340 \\
9 & Unclear cases on visual check & 87 & 186 & 229 \\
\hline
10 & Corrected count of genuine period changes\tablefootmark{c} & 1000 & 534 & 421 \\
11 & Corrected count of wrong periods\tablefootmark{d} & 50 & 106 & 758 \\
12 & Corrected count of unclear cases\tablefootmark{e} & 364 & 2164 & 1512 \\
13 & Corrected count of valid objects\tablefootmark{f} & 2430 & 2803 & 3053 \\
\hline
14 & Corrected period change percentage\tablefootmark{g} & 41.2\% & 19.1\% & 13.8\% \\
\hline
\end{tabular}
\tablefoot{
\tablefoottext{a}{See \citet{lohr14b} for explanation of this ratio.}
\tablefoottext{b}{i.e. row 1 $-$ row 3.}
\tablefoottext{c}{Obtained by extending the confirmed period change ratio (row 6 / row 5) to all objects with plausibly-significant $\chi^2$ ratios (row 3).}
\tablefoottext{d}{Obtained by extending the detected wrong period ratio (row 8 / row 5) to all objects with plausibly-significant $\chi^2$ ratios (row 3).}
\tablefoottext{e}{Obtained by extending the unclear cases ratio (row 9 / row 5) to all objects considered (row 1).}
\tablefoottext{f}{i.e. total objects considered minus expected objects with wrong periods or unclear cases (row 1 $-$ (row 11 + row 12)).}
\tablefoottext{g}{i.e. (row 10 / row 13) $\times$ 100\%.}
}
\end{table*}

There are two plausible causes for such widespread sinusoidal period
variations.  The Applegate mechanism \citep{applegate} could produce
semi-sinusoidal modulations, of amplitude $\Delta P/P\sim10^{-5}$ on a
time scale of decades, in close binaries containing at least one
active, convective star; luminosity variations would also be expected
to be observed with the same period as the O$-$C modulation.  However,
it is unlikely or impossible that this mechanism is responsible for
many of the cases seen here, which include widely-separated
long-period binaries as well as W~UMa-type systems, systems not
exhibiting obvious luminosity changes of the correct period
(e.g. J165649, Fig.~\ref{J165649}), systems exhibiting modulation on
quite short time scales (e.g. J161253, Fig.~\ref{J161253}) and O$-$C
amplitudes substantially too large (e.g. J051927).  The similar shapes
of the three distributions -- varying primarily in scale -- also
suggest a common underlying cause, rather than different mechanisms
operating in different types of system.

A more straightforward explanation, which would be applicable to
nearly all types of eclipsing systems seen here, would be the
influence of a third body inducing sinusoidal period modulation
through the Roemer (light travel time) delay and/or the physical delay
for a third star in an eccentric or inclined-plane orbit
(\citet{rappaport} gives further details of the expected contributions
of each effect in systems of different configurations).  Further
support for this cause is provided by the greater frequency of period
changes seen in short-period EW-type systems compared with long-period
EBs and EAs (Fig.~\ref{perchgall}): the modulation amplitude may be
expected to be greater, and the modulation period shorter, in closer
systems containing low-mass binaries.  Moreover, a third star may have
actually driven a binary to shorter orbital periods and towards
contact configuration through Kozai cycles \citep{kozai}.
Additionally, higher-order multiplicity in some eclipsing SuperWASP
binaries has been strongly supported by other techniques
\citep{lohr13b,lohr13}, so such systems are not inherently unlikely.

Estimates of the modulating periods of our samples of objects whose
O$-$C diagrams show at least one full cycle of sinusoidal period
variation allow us to find approximate ratios of outer to inner
orbits, on the assumption that third bodies are present ($P_L/P_S$ in
terms of \citet{tokovinin2}).  The outer orbits have periods between
about two and seven years (the upper limit being provided by the time
span of SuperWASP archival light curves), giving ratios between 250
and 10000.  These are all well above the minimum ratio for dynamical
stability of triple systems, given by \citeauthor{tokovinin2} as 4.7,
or 47 for systems with high eccentricity of the outer orbit.  As is
shown in \citeauthor{tokovinin2}'s Fig.~7 \citeyearpar{tokovinin2},
``there is no typical or preferred period ratio [...] all allowed
combinations of periods actually happen'', and ratios between
$P_L/P_S$ of five and $10^8$ are found.  There is perhaps a lack of
systems with $P_L<10^3$~d in that figure, but \citet{conroy} have
recently identified 236 candidate Kepler close binaries in triple
systems, 35 of which have $P_L<700$~d.  Thus our visibly
sinusoidally-varying objects would also be fully consistent with the
known period ratios in triples.

In spite of the symmetry of our period change distributions, a few of
the objects here exhibiting quadratic variations in their O$-$C
diagrams might still be better explained by other factors such as mass
transfer or loss, as argued for a selection of 18 known Algols in
\citet{erdem}.  (We may note that two of their cases exhibited very
dramatic period increases of 18.8 and 22.9~s~yr\textsuperscript{-1}:
far greater than any observed here.)  There is, however, no reason why
a system should not exhibit both mass transfer/loss and third body
effects in a single light curve, and \citet{soydugan} claim precisely
this for the Algol-type systems S~Equ and AB~Cas.

If all the period changes measured here were actually associated with
third bodies, their frequency within the sample of SuperWASP eclipsing
binary candidates would allow us to place a lower limit on the
frequency of triples amongst binaries more generally, as around 24\%.
(Of course, some of these detected period changes probably have other
causes, so in a sense it is also an upper limit.)  This value lies
between \citeauthor{tokovinin1}'s slightly higher figure of 29\% for F
and G dwarfs (taking into account detection biases), and
\citeauthor{rappaport}'s estimate of ``at least 20\% of all close
binaries''.

\section{Conclusion}

A neural-net-based catalogue of $\sim$14000 candidate SuperWASP
eclipsing binaries was searched to check their orbital periods and
classification, and to search for evidence of period change.  Numerous
clear cases of quadratic and sinusoidal variation in O$-$C diagrams
were observed; interpreting the quadratic variation as sections of
longer-period sinusoidal variation would explain the symmetrical
period change distributions observed in all three classes of binaries.
If this period modulation is caused by third bodies, this allows us to
estimate a lower limit for the higher-order multiplicity fraction
among local galactic binaries of around 24\%, which tallies well with
other estimates.  In the future, we would hope to confirm some of these
candidate triple systems by direct imaging and/or spectroscopic
follow-up observations.

\begin{acknowledgements}
  The WASP project is currently funded and operated by Warwick
  University and Keele University, and was originally set up by
  Queen's University Belfast, the Universities of Keele, St. Andrews
  and Leicester, the Open University, the Isaac Newton Group, the
  Instituto de Astrofisica de Canarias, the South African Astronomical
  Observatory and by STFC.  This work was supported by the Science and
  Technology Funding Council and the Open University.
\end{acknowledgements}

\bibliographystyle{aa}
\bibliography{reflist}

\Online
\begin{appendix}
\section{SuperWASP eclipsing binaries exhibiting period changes}

\longtab{1}{
\begin{longtable}{l l l l l l l}
\caption{Periods and quadratic period change determinations for 388 EW-type eclipsing binary candidates.  \label{ewquadtab}}\\
\hline\hline
WASP ID (1SWASP & $P$ & $\delta P$ & $\dot{P}$ & $\delta \dot{P}$ & Linear fit & Quadratic fit\\
Jhhmmss.ss$\pm$ddmmss.s) & /s & /s & /s~yr\textsuperscript{-1} & /s~yr\textsuperscript{-1} & reduced $\chi^2$ & reduced $\chi^2$ \\
\hline
\endfirsthead
\caption[]{(continued)}\\
\hline\hline
WASP ID (1SWASP & $P$ & $\delta P$ & $\dot{P}$ & $\delta \dot{P}$ & Linear fit & Quadratic fit\\
Jhhmmss.ss$\pm$ddmmss.s) & /s & /s & /s~yr\textsuperscript{-1} & /s~yr\textsuperscript{-1} & reduced $\chi^2$ & reduced $\chi^2$ \\
\hline
\endhead
J000139.04-362125.6 & 23055.9531 & 0.0035 &  0.10153 &  0.01084 & 1.41 & 1.02\\
J000309.18-345618.4 & 22931.8594 & 0.0024 & -0.03004 &  0.00312 & 1.41 & 1.00\\
J001259.92+141858.3 & 28468.7715 & 0.0047 & -0.16307 &  0.01053 & 2.37 & 1.08\\
J001536.38+312111.8\footnote{This ``object'' and its neighbour(s) at similar coordinates are duplicates of a single astrophysical object.} & 26700.2773 & 0.0050 & -0.11666 &  0.01319 & 1.40 & 1.01\\
J001537.75+312045.4$^1$ & 26700.2617 & 0.0050 & -0.11709 &  0.01377 & 1.43 & 1.05\\
J002328.00-204150.0 & 35829.8828 & 0.0073 & -0.21826 &  0.00243 & 30.75 & 7.23\\
J002446.47+192503.1 & 34684.7734 & 0.0036 & -0.06151 &  0.00638 & 1.70 & 1.02\\
J002449.36-274418.9 & 27100.4863 & 0.0029 &  0.10369 &  0.00627 & 3.01 & 1.39\\
\end{longtable}
}

\longtab{2}{
\begin{longtable}{l l l l l}
\caption{Periods for 92 EW-type eclipsing binary candidates exhibiting sinusoidal period change.  \label{ewsinetab}}\\
\hline\hline
WASP ID (1SWASP & $P$ & $\delta P$ & Linear fit & Quadratic fit\\
Jhhmmss.ss$\pm$ddmmss.s) & /s & /s & reduced $\chi^2$ & reduced $\chi^2$ \\
\hline
\endfirsthead
\caption[]{(continued)}\\
\hline\hline
WASP ID (1SWASP & $P$ & $\delta P$ & Linear fit & Quadratic fit\\
Jhhmmss.ss$\pm$ddmmss.s) & /s & /s & reduced $\chi^2$ & reduced $\chi^2$ \\
\hline
\endhead
J000326.28-163119.6 & 26964.3004 & 0.0035 & 1.77 & 1.22 \\
J000653.04+164621.7 & 23994.2856 & 0.0027 & 1.93 & 1.27 \\
J001527.97-022821.8 & 21791.2432 & 0.0111 & 1.30 & 1.01 \\
J004403.66-023850.7 & 24174.6406 & 0.0029 & 1.85 & 1.09 \\
J010321.75+023040.2 & 24611.0539 & 0.0107 & 2.22 & 1.07 \\
J011611.78-065936.8 & 28058.0934 & 0.0095 & 1.70 & 1.07 \\
J012124.43+031002.9 & 23051.7087 & 0.0040 & 1.69 & 1.03 \\
J012450.31-324123.7 & 26694.9375 & 0.0022 & 4.23 & 2.52 \\
\end{longtable}
}

\longtab{3}{
\begin{longtable}{l l l l l l l}
\caption{Periods and quadratic period change determinations for 137 EB-type eclipsing binary candidates.  \label{ebquadtab}}\\
\hline\hline
WASP ID (1SWASP & $P$ & $\delta P$ & $\dot{P}$ & $\delta \dot{P}$ & Linear fit & Quadratic fit\\
Jhhmmss.ss$\pm$ddmmss.s) & /s & /s & /s~yr\textsuperscript{-1} & /s~yr\textsuperscript{-1} & reduced $\chi^2$ & reduced $\chi^2$ \\
\hline
\endfirsthead
\caption[]{(continued)}\\
\hline\hline
WASP ID (1SWASP & $P$ & $\delta P$ & $\dot{P}$ & $\delta \dot{P}$ & Linear fit & Quadratic fit\\
Jhhmmss.ss$\pm$ddmmss.s) & /s & /s & /s~yr\textsuperscript{-1} & /s~yr\textsuperscript{-1} & reduced $\chi^2$ & reduced $\chi^2$ \\
\hline
\endhead
J001311.15+053540.1 & 26657.5996 & 0.0060 &  0.22048 &  0.01356 & 2.56 & 1.23\\
J002035.20+400415.8 & 25666.2285 & 0.0053 & -0.09488 &  0.00645 & 2.06 & 1.00\\
J002439.13+245522.7 & 36526.1094 & 0.0046 &  0.02151 &  0.00200 & 1.68 & 1.00\\
J003512.87-325603.5 & 43663.8633 & 0.0072 &  0.09966 &  0.00677 & 3.32 & 2.48\\
J010313.78+352903.7 & 38643.6680 & 0.0049 & -0.05079 &  0.00367 & 2.01 & 1.00\\
J010655.30-194334.5 & 29782.3379 & 0.0042 &  0.04943 &  0.00394 & 1.49 & 1.00\\
J011529.82+045222.6 & 65381.7344 & 0.0332 & -1.31450 &  0.09737 & 1.98 & 1.01\\
J011915.76+301338.9 & 25077.4961 & 0.0037 &  0.02525 &  0.00292 & 1.43 & 1.08\\
\end{longtable}
}

\longtab{4}{
\begin{longtable}{l l l l l}
\caption{Periods for 30 EB-type eclipsing binary candidates exhibiting sinusoidal period change.  \label{ebsinetab}}\\
\hline\hline
WASP ID (1SWASP & $P$ & $\delta P$ & Linear fit & Quadratic fit\\
Jhhmmss.ss$\pm$ddmmss.s) & /s & /s & reduced $\chi^2$ & reduced $\chi^2$ \\
\hline
\endfirsthead
\caption[]{(continued)}\\
\hline\hline
WASP ID (1SWASP & $P$ & $\delta P$ & Linear fit & Quadratic fit\\
Jhhmmss.ss$\pm$ddmmss.s) & /s & /s & reduced $\chi^2$ & reduced $\chi^2$ \\
\hline
\endhead
J004716.03-194143.7 & 42233.1360 & 0.0094 & 1.43 & 2.95 \\
J010006.57-074430.5 & 60863.1903 & 0.0594 & 3.07 & 1.08 \\
J023948.86-261804.7 & 42754.9359 & 0.0067 & 1.76 & 1.19 \\
J043127.71-004352.0 & 40229.2524 & 0.0234 & 1.62 & 1.02 \\
J043536.13-005226.5 & 32505.8435 & 0.0468 & 2.27 & 1.06 \\
J090358.97-213218.5\footnote{This ``object'' and its neighbour(s) at similar coordinates are duplicates of a single astrophysical object.} & 40644.4700 & 0.0095 & 1.34 & 1.03 \\
J090401.71-213238.5$^4$ & 40644.4622 & 0.0103 & 1.52 & 1.02 \\
J091816.12-184818.8 & 24490.6721 & 0.0105 & 2.42 & 1.67 \\
\end{longtable}
}

\longtab{5}{
\begin{longtable}{l l l l l l l}
\caption{Periods and quadratic period change determinations for 172 EA-type eclipsing binary candidates.  \label{eaquadtab}}\\
\hline\hline
WASP ID (1SWASP & $P$ & $\delta P$ & $\dot{P}$ & $\delta \dot{P}$ & Linear fit & Quadratic fit\\
Jhhmmss.ss$\pm$ddmmss.s) & /s & /s & /s~yr\textsuperscript{-1} & /s~yr\textsuperscript{-1} & reduced $\chi^2$ & reduced $\chi^2$ \\
\hline
\endfirsthead
\caption[]{(continued)}\\
\hline\hline
WASP ID (1SWASP & $P$ & $\delta P$ & $\dot{P}$ & $\delta \dot{P}$ & Linear fit & Quadratic fit\\
Jhhmmss.ss$\pm$ddmmss.s) & /s & /s & /s~yr\textsuperscript{-1} & /s~yr\textsuperscript{-1} & reduced $\chi^2$ & reduced $\chi^2$ \\
\hline
\endhead
J002717.28-233642.0 & 132393.9688 & 0.0703 &  0.90936 &  0.04890 & 8.76 & 1.10\\
J003518.16+404547.6 & 149296.3125 & 0.0778 &  1.42926 &  0.09536 & 5.23 & 1.01\\
J004345.09+305619.6\footnote{This ``object'' and its neighbour(s) at similar coordinates are duplicates of a single astrophysical object.} & 128416.8203 & 0.0322 & -0.87129 &  0.03805 & 9.82 & 1.10\\
J004346.91+305603.8$^5$ & 128416.7266 & 0.0336 & -0.82979 &  0.04481 & 6.35 & 1.01\\
J005015.63+310004.9 & 148976.7969 & 0.0755 &  0.25266 &  0.04421 & 1.61 & 1.00\\
J005147.36+264032.1 & 116414.1094 & 0.0649 &  0.28170 &  0.03874 & 2.10 & 1.00\\
J005156.72-044304.5 & 146859.7500 & 0.1080 &  1.11732 &  0.27649 & 1.40 & 1.02\\
J005543.22+153749.0 & 105194.4531 & 0.0804 &  0.27323 &  0.04344 & 1.34 & 1.01\\
\end{longtable}
}

\longtab{6}{
\begin{longtable}{l l l l l}
\caption{Periods for 17 EA-type eclipsing binary candidates exhibiting sinusoidal period change.  \label{easinetab}}\\
\hline\hline
WASP ID (1SWASP & $P$ & $\delta P$ & Linear fit & Quadratic fit\\
Jhhmmss.ss$\pm$ddmmss.s) & /s & /s & reduced $\chi^2$ & reduced $\chi^2$ \\
\hline
\endfirsthead
\caption[]{(continued)}\\
\hline\hline
WASP ID (1SWASP & $P$ & $\delta P$ & Linear fit & Quadratic fit\\
Jhhmmss.ss$\pm$ddmmss.s) & /s & /s & reduced $\chi^2$ & reduced $\chi^2$ \\
\hline
\endhead
J012449.35-074004.9 & 48166.4372 & 0.0106 & 1.73 & 1.00 \\
J065342.87-430911.9 & 231569.1013 & 0.1861 & 1.72 & 1.11 \\
J103820.77+141603.6 & 51847.5434 & 0.0130 & 3.63 & 2.10 \\
J154429.44-150320.4 & 153000.8339 & 0.1163 & 1.69 & 1.09 \\
J154946.55-204150.7 & 31539.3276 & 0.0092 & 2.28 & 1.54 \\
J161253.63-022042.2 & 35162.8162 & 0.0071 & 4.15 & 1.10 \\
J171335.36+304236.0 & 195863.4824 & 0.0602 & 3.36 & 2.04 \\
J173329.02+403315.2 & 85462.0986 & 0.0215 & 1.29 & 1.00 \\
J174615.04+452331.8 & 68011.5823 & 0.0091 & 4.37 & 3.21 \\
J182625.19+412014.5 & 223985.1555 & 0.0854 & 1.94 & 1.00 \\
J210421.98+131253.4 & 102913.9002 & 0.0303 & 2.10 & 1.03 \\
J210935.08-225030.2 & 53947.1017 & 0.0128 & 6.08 & 4.78 \\
J213148.77-450242.3 & 162475.5469 & 0.0851 & 3.92 & 2.41 \\
J214614.83+122939.9 & 53129.3740 & 0.0076 & 1.46 & 1.00 \\
J220643.31-434358.4 & 158330.0850 & 0.1138 & 1.67 & 1.01 \\
J224452.75-420045.6 & 81623.9483 & 0.0213 & 2.22 & 1.16 \\
J235755.28-415543.1 & 121213.2071 & 0.0462 & 1.52 & 1.14 \\
\end{longtable}
}

\end{appendix}
\end{document}